\documentclass[12pt]{article}

\usepackage{graphicx}
\usepackage{float}
\usepackage{gensymb}
\usepackage[dvipsnames]{xcolor}
\usepackage{graphicx}
\usepackage{siunitx}
\begin{document}
	\begin{center}
		
		{\large\bf Spontaneous Formation of Double Emulsions at Particle-Laden Interfaces}
		
		Parisa Bazazi and S. Hossein Hejazi
		
		{\it Department of Chemical and Petroleum Engineering, University of Calgary, Calgary, AB T2N 1N4,
			Canada}
	\end{center}
	Double emulsions, due to their compartmental structures, are essential in food, agricultural, and pharmaceutical applications. Traditionally, double emulsifications rely on the presence of both oil-soluble and water-soluble surfactants or external stimuli responsive materials and require sequential droplet formation settings or unique fluidic designs. We report on unusual phenomenon where double emulsions are spontaneously formed as soon an aqueous nanoparticle dispersion is placed in contact with an oleic micellar solution. Nanoscale water droplets nucleate in oil in the form of swollen micelles. Nanoparticles form a water-shell encapsulating the saturated oil phase with swollen micelles over time. Remarkably, we find that the gradual surface-activation of nanoparticles is key in self-double emulsification and controlling the emulsion intensity. We build on this new discovery and design a novel system for double emulsion formation. This approach is a scalable self-sequential strategy for preparing core-shell double emulsions that disperses nanoparticles in the opposite phase by employing micelles as transport vehicles. Incorporating nanoparticles into spontaneous emulsification systems opens novel routes for designing emulsion-based materials.
	\newpage
	\textbf{}\\
	\textbf{INTRODUCTION}\\
	Double emulsions, in which liquid droplets are encapsulated inside secondary droplets, have great potential in a broad class of applications ranging from foods, cosmetics, pharmaceuticals, and drug delivery to agriculture[1-5]. There are numerous one-step and two-step techniques for double emulsification, including phase separation, phase inversion, and microfluidic techniques [6-9]. Emulsion phase inversion and phase separation often constrain the scope of suitable materials [9-11].Microfluidic techniques require unique design and fabrication procedures [6, 12-14]. The significant challenges facing the current methods are reducing the droplet size to the nanoscale, improving production efficiency, and minimizing the energy required for emulsification [7,8,15]. Thus, developing new double emulsification routes to broaden the range of suitable materials, improve droplet stability, and enhance the scale of production without requiring complicated designs and high energy inputs is of great technological importance.
	
	Spontaneous emulsification is an easy and reproducible approach for generating submicrometer droplets without inputting energy  [16,17]. TTherefore, it could be an alternative approach for forming double emulsions with nanometer droplet sizes. Micelles, which are molecular entities comprising surfactants in a liquid medium, are critical to spontaneous emulsification in systems with considerable interfacial tension [18,20]. The preparation of double emulsions requires the sequential formation of water in oil (W/O) and oil in water (O/W) emulsions or vice versa [8]. Therefore, the existence of both water-soluble and oil-soluble surfactants is indispensable [8,21]. However, the simultaneous use of oil-soluble and water-soluble surfactants results in the formation of either (W/O) or (O/W), depending on the hydrophilic-hydrophobic balance (HLB) of the surfactant mixture, and does not result in the formation of a double emulsion [22,23]. To formulate double emulsions, surfactants should be added sequentially to the oil and water phases [8]. Considering the slower adsorption rate of colloidal particles to oil-water interfaces compared to that of commercial surfactants, they could be a promising option for generating double emulsions [24]. 
	
	Particles are used to form double emulsions [25-27] and  to improve the stability of droplets as they create a mechanical barrier at the oil-water interface [7,11,28-31]. Multiple emulsions can be generated in a single step by using a homogenizer or an ultrasonicator when the pH or temperature-responsive nanoparticles are pre-dispersed in the oil phase and water is subsequently added. In such cases, the surface chemistry of the particles changes by varying the temperature and pH; hence, the emulsion undergoes phase inversion, resulting in double emulsion formation [25,26]. In another study, multiple emulsions have been successfully formed through phase inversion with fumed silica nanoparticles dispersed in the oil phase (triglyceride or PDMS oils) at concentrations above 1 wt.\% [27]. In summary, multiple emulsions are formed when particles are pre-dispersed in the oil phase and in high particle concentrations where particles are in the aggregated form. However, there have been no studies regarding the spontaneous formation of double emulsions using nanoparticles and micelles. The incorporation of colloidal particles in spontaneous emulsification has two characteristics that will facilitate double emulsification. First, a new surface-active component is generated from the colloidal particles over time. Second, the micelles can encapsulate and disperse colloidal particles in a surrounding medium, regardless of nanoparticle hydrophobicity and liquid polarity.
	
	In this work, we develop a novel and reproducible approach for self-double emulsification by harnessing nanoparticles to invert the initial curvature of the oil-water interface. In this regard, the oil phase contains a high concentration of Span 80 micelles, and silica nanoparticles are uniformly dispersed in the aqueous phase. In-situ emulsification is triggered by micelles impinging on the oil-water interface and adsorbing water and nanoparticles, which results in the formation of nanoscale W/O emulsions. Simultaneously, silica nanoparticles adsorb surfactants from the oil-water interface and/or from micelles dispersed in the water phase. Consequently, the surface-activated nanoparticles migrate to the interface, change the initial curvature of the interface, and generating the second emulsion from the initial W/O emulsion. Scanning electron microscopy verifies the structure of the W/O emulsion which remains intact upon adding nanoparticles. Dynamic light scattering shows that the average size and polydispersity of the droplets shifted slightly over time. The intensity of the double emulsion strongly depends on the concentration of nanoparticles, where increasing the concentration above a specific value lessens the intensity of the double emulsification. We discuss the principle of double emulsification based on the physicochemical properties of the nanoparticle-micelle interactions in great detail. Previous reports on spontaneous double emulsion formation have been limited to block copolymers. The presence of salt impurities in amphiphilic block copolymers results in osmotically driven stress between the internal and external water phases, generating small water droplets at the oil-water interfaces  [15,33-34]. Additionally, in these studies, O/W emulsions are first generated by ultrasonication or homogenization. Later, W/O emulsions are spontaneously formed at the surface of the oil droplets. We believe our developed approach for self-double emulsification is conceptually simple, scalable, and applicable to a wide range of applications, including the encapsulation of nanoparticles in emulsion droplets and surrounding liquid mediums.
	
	\textbf{}\\
	\textbf{RESULTS}\\
	\textbf{Effect of nanoparticles on the spontaneous emulsification systems.} To study the role of nanoparticles in in-situ emulsifications, we gently placed DI water and 4.0 wt.\% silica nanoparticle dispersions in contact with 1.0 and 5.0 wt.\% solutions of Span micelles. In all systems, a white zone is formed in the vicinity of the oil-water interface, indicating spontaneous emulsion formation. The oil phase remains transparent when in contact with the DI water. However, it becomes cloudy in the presence of nanoparticles, and this cloudy zone grows over time (\textbf{Figure 1a} and \textbf{Figure S1a-b}). The emulsion droplets grow as a function of time. The droplets are in the range of 100-250 nm; thus, the solution is a nanoemulsion. The emulsion droplets generated from the nanoparticle dispersion are slightly larger than those generated from DI water (\textbf{Figure 1b}). The polydispersity index (PDI) of the droplets remains constant during the first 150 minutes. However, after this time, the PDI increases to two- to four times its initial value in the absence and presence of nanoparticles, respectively (\textbf{Figure 1c}). The increase in the PDI values indicates the presence of larger droplets.
	
	The cryo-scanning electron microscopy (cryo-SEM) images of the emulsions 180 minutes after initial contact between the oil and aqueous phases show that the oil phase (dark color) is fully covered with small water droplets (light color). The droplets are 200-nm in size, which is consistent with the size indicated by DLS (\textbf{Figure 1d}). The observed water droplets are reversed micelles that are with the DI water and nanoparticle dispersion. The micelles remain spherical in the presence of nanoparticles. To verify the presence of silica nanoparticles in the emulsion phase, we measured the density of the oil and aqueous phases before and after bring the two phases into contact. The density of the silica dispersion does not increase upon contact with the micellar solution (\textbf{Figure S7e}). Additionally, the densities of the emulsions generated from the nanoparticle dispersions are higher than those of the emulsions generated from DI water (\textbf{Figure 2d} left axis). The constant density of the nanoparticle dispersions before and after contact with the micellar solutions and the higher density of emulsions generated from nanoparticle dispersions confirm the presence of silica in the emulsions.
	
	Surprisingly, we observe double emulsion droplets in 5.0 wt.\% Span micellar solutions in contact with both DI-water and nanoparticle dispersion and 1.0 wt.\% Span micellar solution in contact with the nanoparticle dispersion (\textbf{Figure 1d-e}). Double emulsions have been previously formed when both oil-soluble and water-soluble surfactants are present 8. However, our experiments unexpectedly revealed the spontaneous generation of double emulsions in systems containing only oil-soluble surfactants. Increasing the Span concentration from 1.0 to 5.0 wt.\% results in the formation of double emulsions. The presence of silica nanoparticles in the water phase intensifies the double emulsion formation. The generated multiple emulsions have a unique structure: oil drops (with the radius of $\sim 50$ \si{\micro\metre}) are fully covered with nanoscale water droplets while it is encapsulated in an exterior water shell (\textbf{Figure 1e}).
	
	\begin{figure}[H]
		\centering
		\includegraphics[scale=0.45]{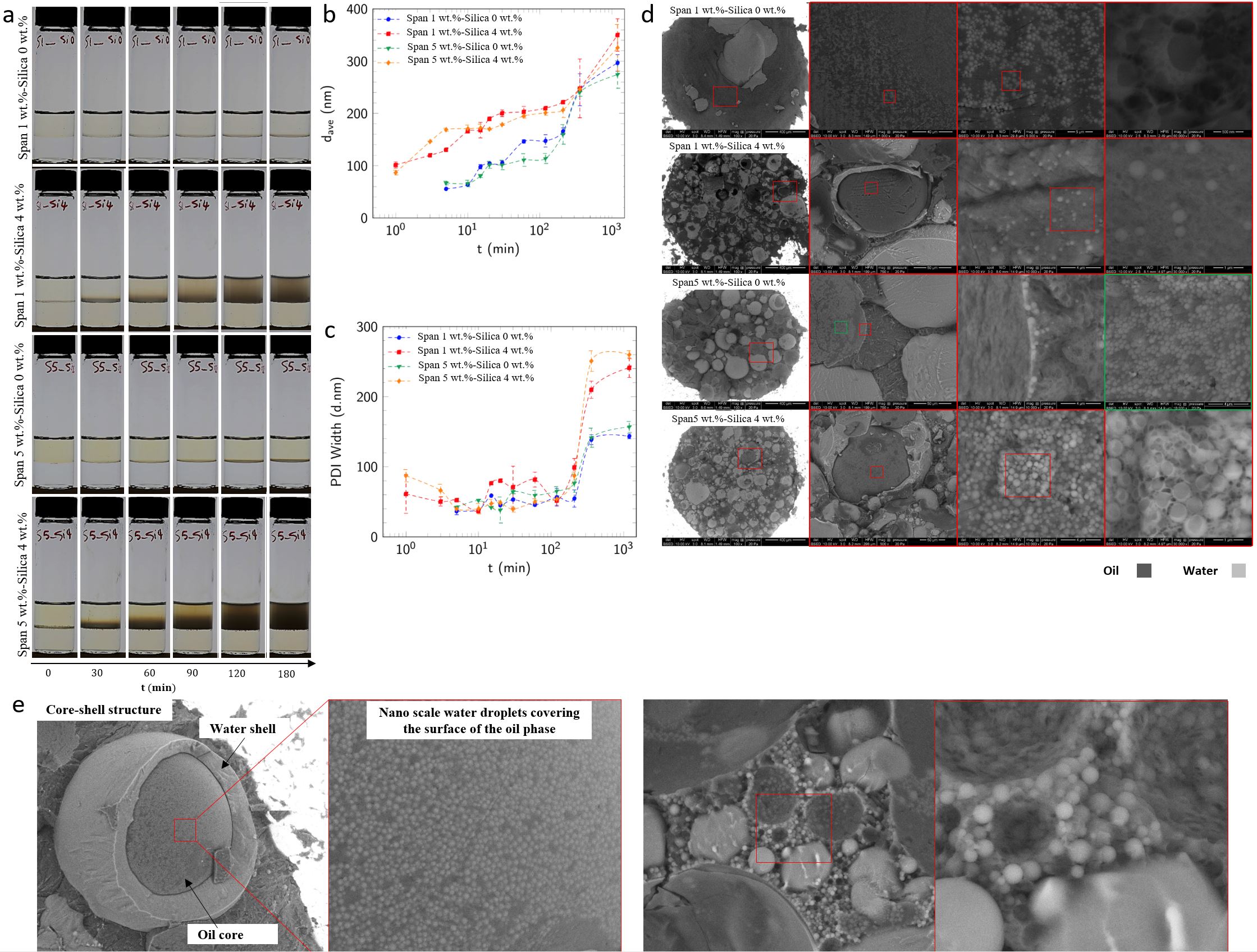}
		\caption{Spontaneous formation of emulsions in micellar solutions-nanoparticle dispersions systems. (\textbf{a}) The rows show emulsion formation with Span 1.0 wt.\% - silica 0.0 wt.\%, Span 1.0 wt.\% - silica 4.0 wt.\%, Span 5.0 wt.\% - silica 0.0 wt.\%, and Span 5.0 wt.\% - silica 4.0 wt.\%, respectively, over time. The aqueous phase is poured into the vial, and the oil phase is gently added from the top. Images were taken after 0, 30, 60, 90, 120, and 150 minutes of micellar solution-aqueous phase contact. (\textbf{b}) Evolution of the average size of the reversed micelles in Span 1.0 wt.\% - silica 0.0 wt.\%, Span 1.0 wt.\% - silica 4.0 wt.\%, Span 5.0 wt.\% - silica 0.0 wt.\%, and Span 5.0 wt.\% - silica 4.0 wt.\% systems; the data are shown in blue, red, green, and orange, respectively. (\textbf{c}) Evolution of the polydispersity index (PDI) of the reversed micelles. (\textbf{d}) Cryo-scanning electron microscopy (Cryo-SEM) micrograph of the generated emulsions. The dark color indicates the oil phase, and the light color represents the aqueous phase. (\textbf{e}) Cryo-SEM micrograph of the double emulsions.}
	\end{figure}
	
	\textbf{}\\
	\textbf{Influence of nanoparticle and micelle concentrations on the double emulsion intensity.} We took images of the emulsification zone as the silica concentration was varied from 0.0 to 40.0 wt.\%, and the Span concentration was changed from 1.0 to 10.0 wt.\%. The images are presented in the plane of silica-Span concentrations, and the zones of the single and double emulsions are indicated (\textbf{Figure 2a}). The oil, silica particles, and water are shown in green, red, and black, respectively. Image quantification revealed that the double emulsion intensity reaches a maximum at a nanoparticle concentration of 4.0 wt.\% in 1.0, 2.5 and 5.0 wt.\% Span micellar solutions. The maximum intensity with 10.0 wt.\% Span micellar solution was achieved with the 10.0 wt.\% silica nanoparticle dispersion  (\textbf{Figure 2b}). The Density and water content (calculated as (($\rho \textsubscript{emulsion}-\rho \textsubscript{micellar solution}$)/($\rho \textsubscript{aqueous phase}-\rho \textsubscript{micellar solution}$))$\times $100) of the emulsions reach the maximum at 4.0 wt.\% Silica concentration (\textbf{Figure 2d} left and right axis, respectively), which further confirms that the maximum emulsification was achieved at this silica concentration.
	
	In all cases, the double emulsion intensity is considerably reduced at a silica nanoparticle concentration of 40.0 wt.\%. This nonlinear dependency of emulsion intensity on the silica concentration can be the particle size effect. The average size of the silica nanoparticles in the 1.0, 2.0, and 4.0 wt.\% dispersion is ${7.9 \pm 0.1\  nm}$, while it increases to ${194.2 \pm 2.7\  nm}$ in the 40.0 wt.\% dispersion (\textbf{Figure S7b}). The initial size of the reverse micelles in Span 80 micellar solutions is $4.1\pm0.1$ nm (\textbf{Figure S7c}). We speculate that since the initial size of the reversed micelles is much smaller than the aggregated particles in the 40.0 wt.\% dispersion, the micelles cannot take up the silica nanoparticles. Thus, the 40.0 wt.\% silica dispersion generates less emulsion compared to the other tested silica concentrations (\textbf{Figure S1a-b} and \textbf{Figure 2a-b}). We also used a 2.5 wt.\% silica nanoparticle dispersion with an average particle size of 100 nm in contact with Span 10 wt.\% to confirm the effect of particle size on the double emulsion intensity. Similar to the 40.0 wt.\% silica dispersion, the intensity of the double emulsion is less than 20\%. The reduced double emulsion intensity with 40.0 wt.\% silica dispersion is discussed further in the following section “mechanism of double emulsion formation”.
	
	We took images of the emulsion zones every 60 minutes after the initial contact between the oil and aqueous phases, and the double emulsions were generated after 180 minutes. The considerable increase in the PDI values after 150 minutes could be an indicator of double emulsion formation (\textbf{Figure 1c}). The generated double emulsions were observed to be stable for 7 days, and their intensity either increases (1.0 wt.\% Span 80 micellar solution) or remains constant (5.0 wt.\% Span 80 micellar solution) from day 1 to day 7 (\textbf{Figure 2c}).
	
	\begin{figure}[H]
		\centering
		\includegraphics[scale=0.53]{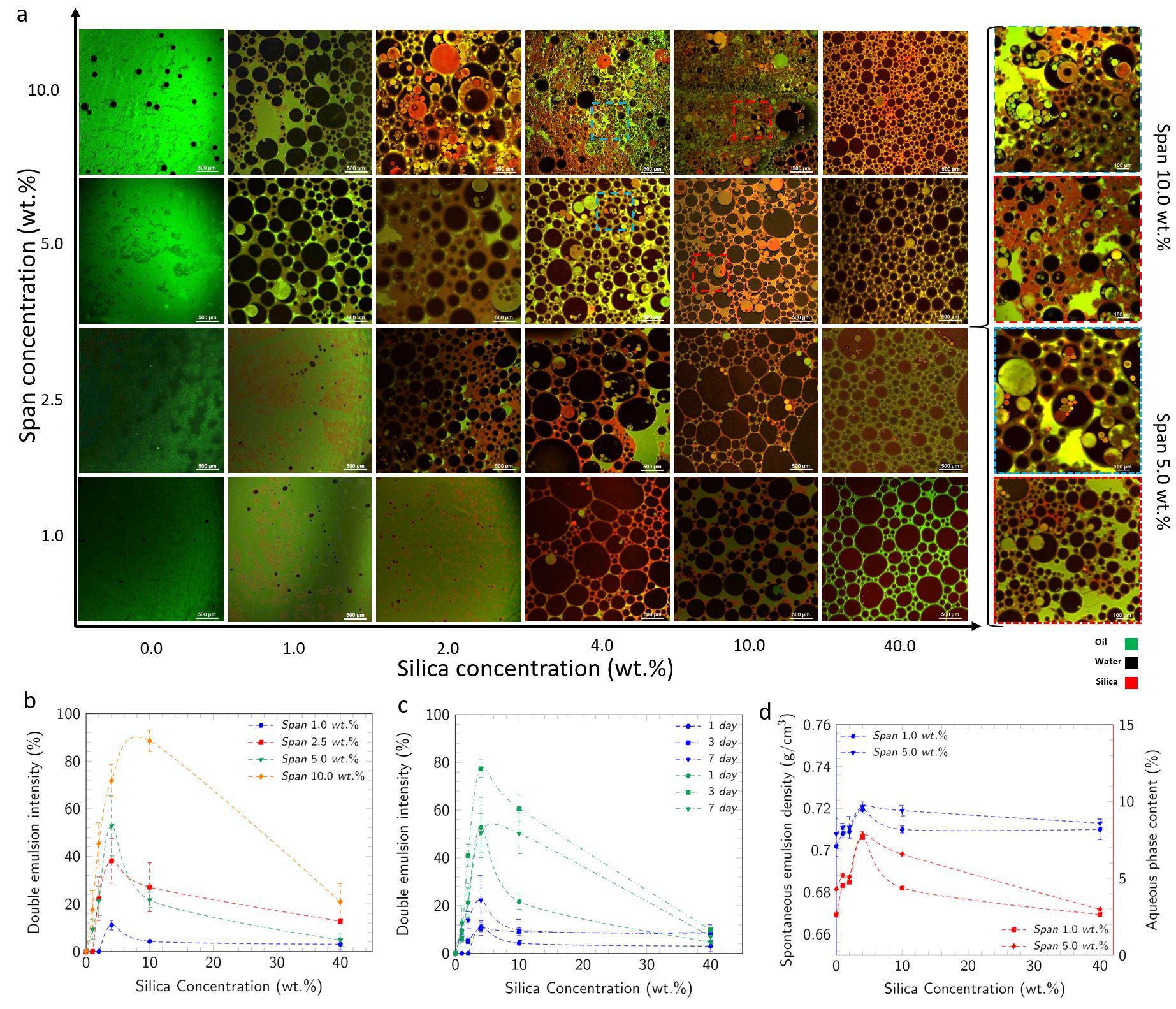}
		\caption{Double emulsion intensity. \textbf{(a)} Double emulsion formation maps obtained with 4X magnification confocal microscopy showing the spontaneous emulsions generated with 1.0, 2.5, 5.0, and 10.0 wt.\% Span concentrations and 0.0, 1.0, 2.0, 4.0, 10.0, and 40.0 wt.\% silica concentrations. The images on the right are magnified (10X magnification) images from the indicated areas on the maps. Oil, nanoparticles, and water are shown in green, red, and black, respectively. The images were captured 1 day after oil-aqueous phase contact. \textbf{(b)} Double emulsion intensity (number of double emulsion droplets/total number of droplets) as a function of silica nanoparticle concentration with 1.0, 2.5, 5.0, and 10 wt.\% Span concentrations are shown in blue, red, green, and orange, respectively. \textbf{(c)} Double emulsion intensity as a function of silica nanoparticle concentration for 1.0 and 5.0 wt.\% Span concentrations after 1 (diamond marks), 3 (square marks), and 7 (inverted triangle marks) days of oil-aqueous phase contact.}
	\end{figure}
	\textbf{}\\
	\textbf{(d)} Density of the emulsion with 1.0 (circle marks) and 5.0 (inverted triangle marks) wt.\% Span concentrations as a function of Silica nanoparticle concentration in left axis (blue) and the aqueous phase content ((($\rho \textsubscript{emulsion}-\rho \textsubscript{micellar solution}$)/($\rho \textsubscript{aqueous phase}-\rho \textsubscript{micellar solution}$))$\times $100) in the emulsion of 1.0 (square) and 5.0 (diamond) wt.\% in right axis (red). To calculate water content, we assume that Silica nanoparticles diffuse uniformly to the emulsion phase. 
	
	\textbf{}\\
	\textbf{Mechanisms of double emulsion formation.} The presence of both oil-soluble and water-soluble surfactants is sought to be required for double emulsification. Since in our experiments, we only used oil-soluble surfactants, we expect the spontaneous production of a water-soluble surface-active material. To confirm this hypothesis, we put 3 ml of DI water and 4.0 wt.\% silica dispersion in contact with 3 ml of pure heptane for 180 minutes. We separated the aqueous phase and measured the surface tension (ST). The ST values for both the DI water and silica dispersion were $71.0\pm0.5$ mN/m, and it does not change over time. This result confirms the absence of surface-active materials in the aqueous phase before and after being in contact with pure heptane. We conducted similar experiments with 1.0 and 5.0 wt.\% Span micellar solutions. The ST of the DI water after contact with the two micellar solutions decreased from 71 to 68 mN/m, and these values remained constant over time. However, the ST of the silica dispersion also decreases from 71 to 57 and 53 mN/m after contact with 1.0 and 5.0 wt.\% micellar solutions, respectively \textbf{(Figure 3a}). The considerable reduction in the ST of silica dispersions indicates the generation of surface-active materials in the aqueous phase. Upon migration into the water phase, Span molecules can attach to the silica nanoparticles and form a surface-active colloid. Particle size analysis of the DI water after contact with the 5.0 wt.\% Span solution also showed the presence of objects with the same average size as that of the micelles (4 nm) (\textbf{Figure 3b}). We speculate that these objects are the Span micelles that migrated into the aqueous phase.
	
	We propose two mechanisms for spontaneous double emulsion formation, as presented in \textbf{Figure 3c} and \textbf{Figure 3d}. In the first mechanism, double emulsions are generated due to surface-activation of the silica nanoparticles. The surfactants (micelles) are initially in the oil phase, the oil-water interfacial curvature is concave toward the oil phase, and the formation of a W/O emulsion is expected. Span is a nonionic surfactant that is also partially soluble in water. Therefore, Span molecules can diffuse into the aqueous phase, especially when the oil phase is oversaturated, and the micelles are aggregated (\textbf{Figure S7d}). Once the surfactants comes into contact with the silica nanoparticles, either at the interface or in the bulk, they interact with the particles and generate water-soluble surface-active materials. Thus, the surface-activate particles in the water migrate to the interface, change the curvature of the interface, and result in the formation of O/W emulsions from the initial W/O emulsion (\textbf{Figure 3c}). 
	
	In the \textbf{second mechanism}, double emulsions are formed due to the hydrophobic interactions between the tails of the Span molecules. This mechanism only occurs in oversaturated Span micellar solutions (\textbf{Figure S7d}). Reverse micelles remove water molecules from the aqueous phase by directly impinging on the interface. The swollen reverse micelles act as solutes and disperse in the oil phase by osmosis. At high Span concentrations, the micellar solution is oversaturated (5.0 wt.\%), and surfactants attach their hydrophobic tails to the tails of other surfactant molecules in the micelle to minimize their contact with the oil phase. This arrangement generates a chain of hydrophilic heads around the filled micelles. These hydrophilic heads adsorb water molecules and create a water shell (\textbf{Figure 3d}). We expect both mechanisms to be simultaneously active in our systems.  
	
	
	\begin{figure}[H]
		\centering
		\includegraphics[scale=0.45]{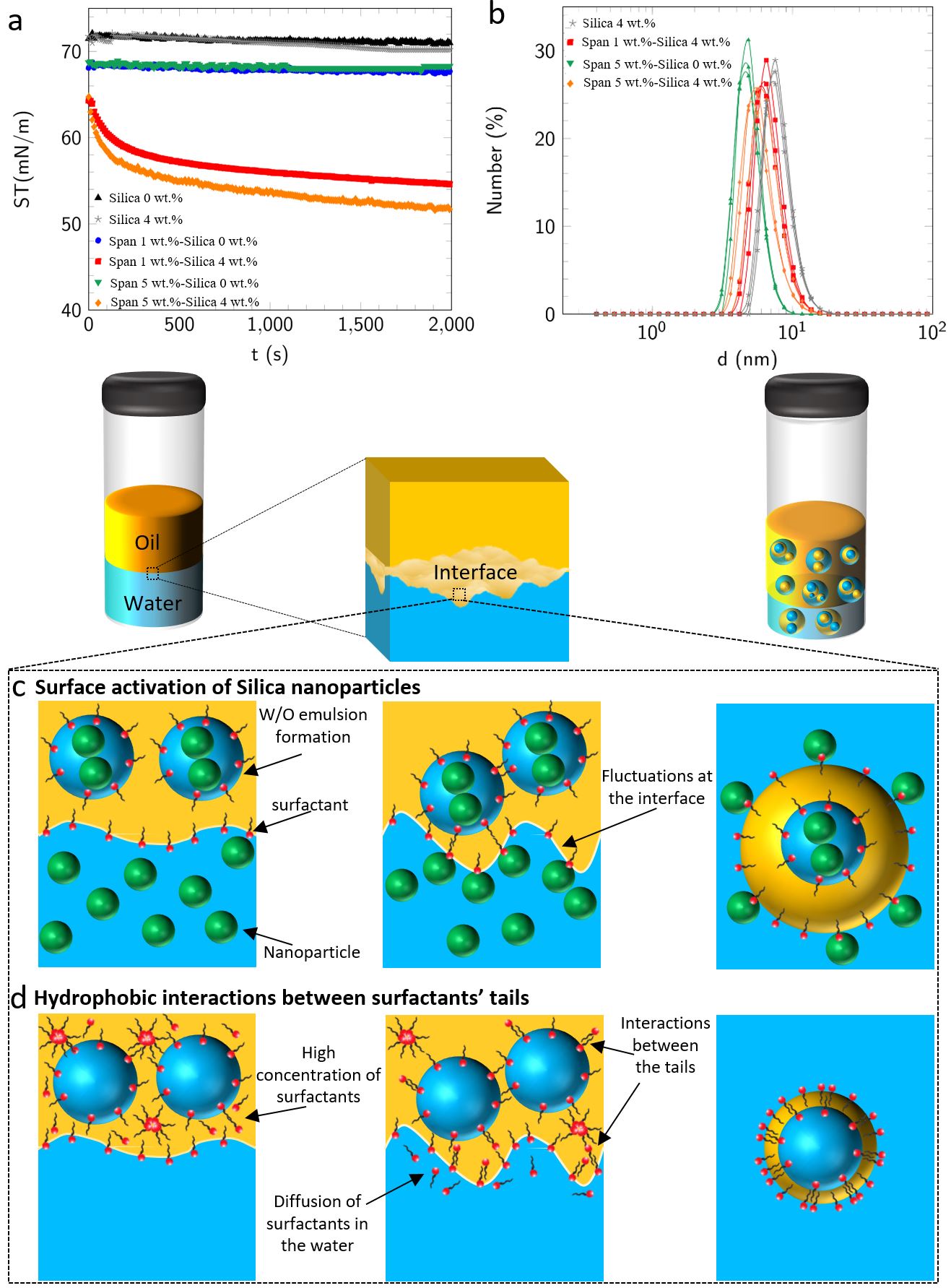}
		\caption{Mechanism of double emulsion formation. \textbf{(a)} Surface tension (ST) of the aqueous phase after 180 minutes being in contact with the micellar solution of Span 1.0 wt.\% - Silica 0.0 wt.\%, Span 1.0 wt.\% - Silica 4.0 wt.\% , Span 5.0 wt.\% - Silica 0.0 wt.\%, and Span 5.0 wt.\% - Silica 4.0 wt.\%  systems in blue, red, green, and orange, respectively. The ST of water and 4.0 wt.\% silica dispersion after 180 minutes being in contact with heptane is plotted in black and gray for the sake of verification. \textbf{(b)} Particle size distribution of the aqueous phase after 180 minutes being in contact with the oil phase for  Span 1.0 wt.\% - Silica 4.0 wt.\% , Span 5.0 wt.\% - Silica 0.0 wt.\%, and Span 5.0 wt.\% - Silica 4.0 wt.\%  systems in red, green, and orange, respectively. \textbf{(c)} Double emulsion formation mechanism 1: Surface activation of Silica nanoparticles. \textbf{(d)} Double emulsion formation mechanism 2: hydrophobic interactions between surfactant tails. In the schematics (c, d) yellow, blue, red, and green are oil, water, surfactant, and nanoparticle, respectively.}
	\end{figure}
	\textbf{}\\
	\textbf{Nanoparticle-micelle (Silica-Span) interactions.} To understand the silica-Span interactions, we measured the dynamic IFT of the micellar solution-nanoparticle dispersions. For all Span concentrations, the IFT decreases when silica nanoparticles are added to the system (\textbf{Figure 4}). At least two factors affect the dynamic IFT of the present systems: (i) the surface-activation of the silica nanoparticles over time and (ii) surfactant removal from the interface due to spontaneous emulsification. In the systems in which the initial IFT (t=0) is above 4.0 mN/m, the IFT monotonically decreases over time. Since the initial IFT is high, emulsification occurs at a much slower rate than nanoparticle activation and migration to the interface. Therefore, the interface is saturated with surface-active materials, and the IFT decreases, reaching a plateau after approximately 10 minutes. In contrast, in systems with a very low initial IFT, less than 0.6 mN/m, IFT shows an increasing trend. In systems which have high concentrations of surfactants and nanoparticles, spontaneous emulsification is a fast process. The generated emulsion droplets continuously remove surface-active materials as they detach from the interface. Hence, the IFT values increase over time. It is noted that the rate at which the IFT increases for the 40.0 wt.\% silica dispersion is lower than those for the other silica dispersions. In fact, the aggregated silica nanoparticles reduce the rate of emulsion formation which  is consistent with the presented data on the double emulsion intensity (\textbf{Figure 2a-b}). When the initial IFT is 1.0-2.5 mN/m, the system shows nonmonotonic behavior. The IFT initially decreases (100-200 minutes) but then increases for the remainder of the measurements. The point at which these trends switch depends on the silica-Span concentrations. We suggest that at this point, the rate of silica migration is comparable to the rate of removal of the surface-active materials by the emulsion droplets.
	
	The lower IFT values for the micellar solution-silica dispersions confirm silica-Span interactions. There are at least two possible interactions between silica and Span: hydrophobic interactions and hydrogen bonding. Hydrophobic interactions occur between the nonpolar tail of the surfactants and the hydrophobic sites on the silica particles [35,36]. Hydrogen bonding occurs between the polar head groups of the surfactants and the particles (between the oxygen atom of the ester group or the hydroxyl group of the Span molecules and protonated hydroxyl groups on the silica[37]. The measured elasticity of the interfacial layers decreases upon addition of silica particles to the aqueous phase. The reduction in the interfacial elasticity confirms spontaneous emulsion formation and their release from the interface during drop oscillation (\textbf{Figure S9f}).
	\begin{figure}[H]
		\centering
		\includegraphics[scale=0.55]{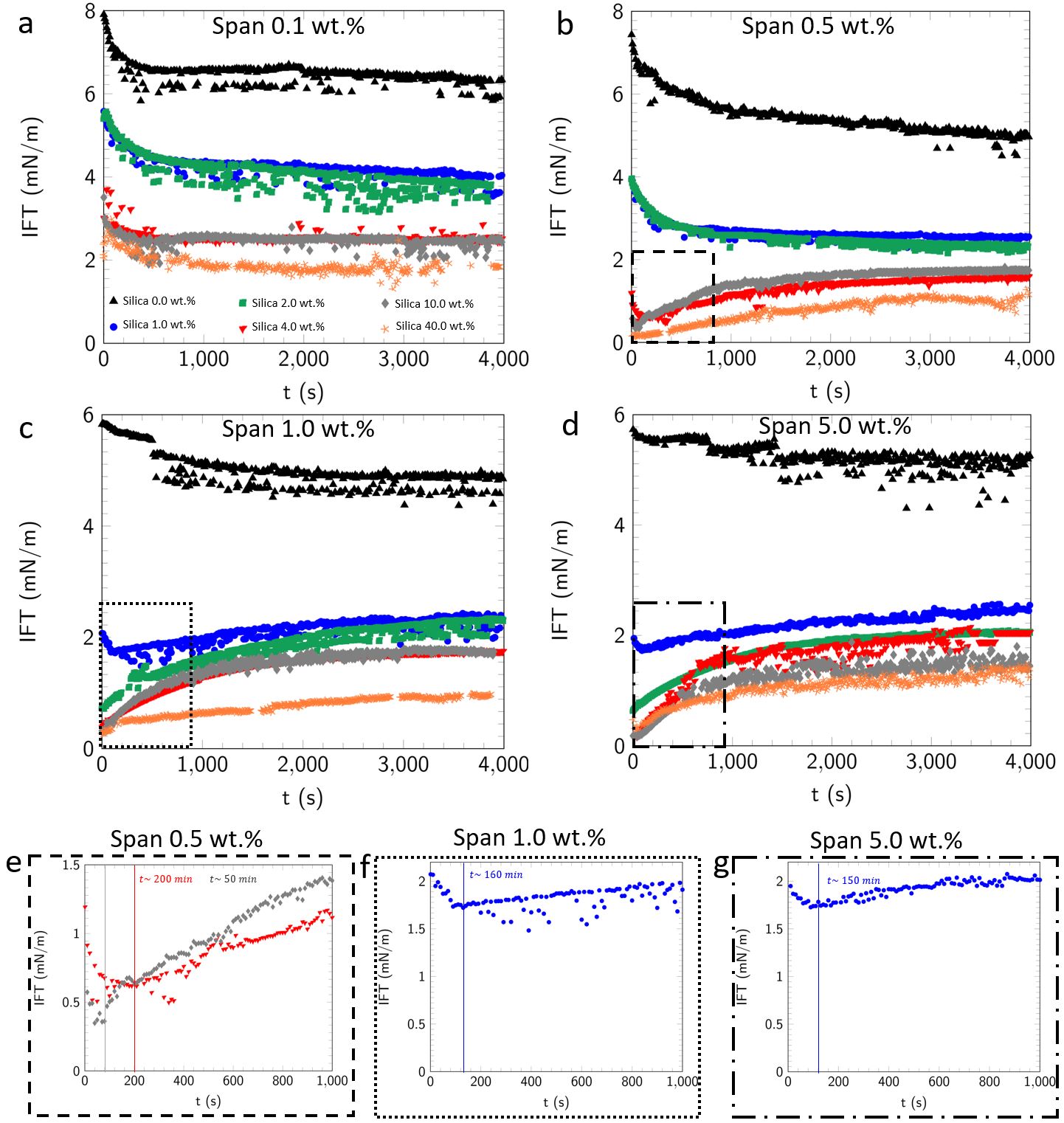}
		\caption{Dynamic interfacial tension (IFT) of micellar solution-Silica dispersionss. Span 80 concentration in the oil phase is fixed in each figure as \textbf{(a)} 0.1 wt.\%, \textbf{(b)} 0.5 wt.\%, \textbf{(c)} 1.0 wt.\%, and \textbf{(d)} 5.0 wt.\% . In each figure black, blue, green, red, gray, and orange represent the data of 0.0, 1.0, 2.0, 4.0, 10.0, and 40.0 wt.\% Silica nanoparticle dispersions. \textbf{(e)}-\textbf{(g)} show the magnified regions of \textbf{(b)}-\textbf{(c)} in dashed, dotted, and dash dotted frames. }
	\end{figure}
	\textbf{}\\
	\textbf{Universality and scalability of the approach.} The gradual surface-activation of the nanoparticles by surfactant adsorption is a well-known physical process, and it is not limited to a specific type of particle [38,39]. Therefore, spontaneous double emulsification can be expected in other surfactant-particle systems as long as there are high concentrations of surfactants and interactions between surfactants and particles. Also, the spontaneous double emulsification is not limited to the use of hydrocarbon oils. We conducted experiments with two samples of biocompatible mineral oils with varying viscosities (30 and 135 mPa.s). Double emulsions were formed in both systems but at different time scales. In contrast to the heptane system in which the double emulsions are formed after a few hours (3 hr), double emulsions are generated after 5 days for the 30 mPa.s oil and after 7 days for 135 mPa.s oil (\textbf{Figure 5}). For all systems, we show the emulsion maps on the plane of silica-Span concentrations, and for these samples, the maximum intensity of the double emulsion was achieved with 10.0 wt.\% silica and 10.0 wt.\% Span and with 4.0 wt.\% silica for lower Span concentrations. The generated double emulsions for the low viscous mineral oil have similar characteristics to those formed with heptane. In high viscous mineral oil systems, especially at 10.0 wt.\% Span concentration, tortuous and interconnected structures of oil and aqueous phases are formed. These bicontinuous structures signal the possibility of the spontaneous formation of bijels, which needs further experimentation and analysis to be verified. 
	
	\begin{figure}[H]
		\centering
		\includegraphics[scale=0.4]{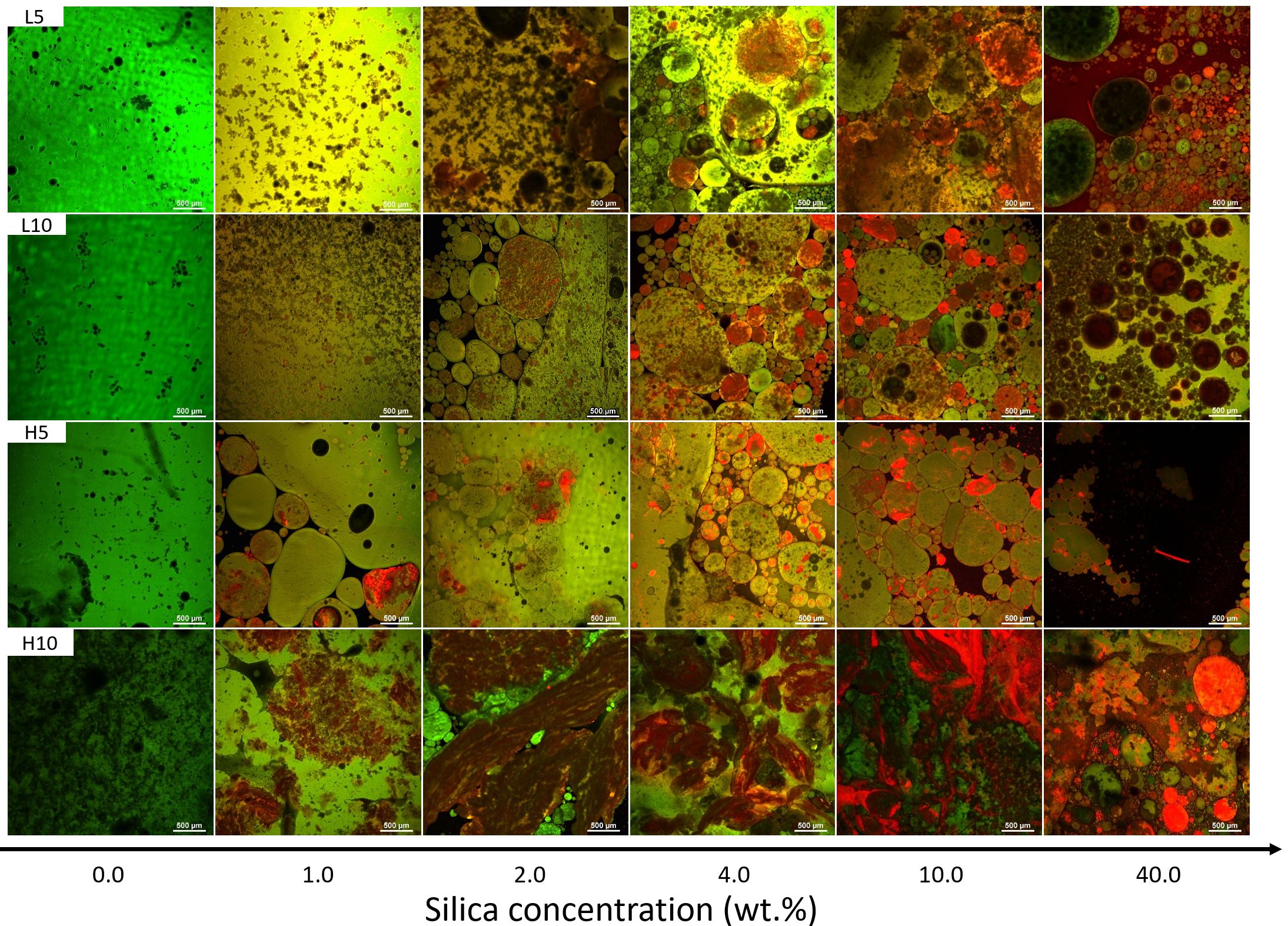}
		\caption{Double emulsion formation map with two mineral oils with $30$ and $135 \ mPa.s$ viscosities names as low and high viscous mineral oils. Samples are prepared with two Span concentrations for each mineral oil (5.0 and 10.0 wt./\%). On the map, L5, L10, H5, H10 are refereed to 5.0 wt.\% Span in low viscous oil, 10.0 wt.\% Span in low viscous oil, 5.0 wt.\% Span 80 in high viscous oil, and 10.0 wt.\% Span in low viscous oil. Oil, particles, and water are shown in green, red, and black, respectively.}
	\end{figure}
	\textbf{}\\
	\textbf{DISCUSSION}\\
	In summary, we demonstrated a novel approach for spontaneously generating double emulsions. The presence of Span micelles in the oil phase results in the formation of W/O nanoemulsions. Silica nanoparticles gradually adsorb surfactants or micelles, generating water-soluble surface-active components, consequently forming O/W emulsions. Since the oil is already a W/O nanoemulsion, the emulsion generated in the second step is a W/O/W double emulsions. Increasing the silica concentration from 1.0 to 4.0 wt.\% and the Span concentration from 1.0 to 10.0 wt.\% increases the intensity of the double emulsion. Micelle concentrations in the Span solution above 5.0 wt.\% result in the formation of a water-soluble material due to the hydrophobic interactions between the tails of the surfactant molecules, leading to double emulsion formation. We reported that silica nanoparticles considerably increase the rate of emulsion formation, and IFT monotonically increases over time. Our approach is not limited to low-viscosity oils, and we documented double emulsion formation with mineral oils with viscosities of 35 and 135 mPa.s. Our double emulsification method with nanoparticles is spontaneous and differs significantly from approaches previously reported in the literature. Previous attempts for spontaneous multiple emulsification are limited to the use of block copolymers, and the emulsification was not spontaneous from the first step. The primary purpose of the incorporation of colloidal particles in double emulsions is to increase the stability of the oil-water interface. A few studies have reported the unexpected formation of double emulsions when silica nanoparticles were dispersed in triglyceride or PDMS oils. Homogenizers generated primary W/O emulsions, and W/O/W emulsions were produced due to phase inversion at high water contents. We report on the unusual phenomenon of the spontaneous formation of double emulsions from the early stage. We expect that our findings on the use of Span-silica for spontaneous double emulsion formation can be extended to other micelle-nanoparticle systems by considering three criteria: (i) the nanoparticles and micelles are predispersed in two different phases, (ii) the nanoparticles are considerably smaller than the swollen micelles, and (iii) the nanoparticles gradually become surface-active. Our discovery allows the straightforward preparation of nanoscale droplets that can spontaneously encapsulate both hydrophilic and hydrophobic cargoes. Tuning the micelle-nanoparticle concentration allows the formation of double emulsions with unique morphologies and provides a multifunctional template for a broad range of applications. 
	\newpage
	\textbf{}\\
	\textbf{METHOD}\\
	\textbf{Materials.} We use a 40 wt.\% silica dispersion (HS 40, Sigma) as the concentrated source of nanoparticles. Silica nanoparticles in the dispersion are spherical with an average diameter of $\SI{7.9 \pm 0.1}\  nm$. We also use a 2.5 wt.\% fluorescent silica dispersion (DNG-L007, Creative Diagnostics), with a diameter of 100 nm. Sorbitan monooleate (Span 80, Sigma) is used as an oil soluble non-ionic surfactant. Span 80 has the molecular weight of $428.60 \  g/mol$, the hydrophilic-lipophilic balance (HLB) of 4.3, and the critical micelle concentration of $1.8 \times 10^{-2} \ mM$ in heptane. We use heptane (anhydrous, 99\%, Sigma) as the model oil. A viscous mineral oil sample (Drakeol 35: $ \rho=0.88 \ g/cm^3$, $ \mu= 135 \ mPas$) and a mineral oil with a lower viscosity (Sigma, $ \rho =0.84\  g/cm^3$, $ \mu= 30 \ mPas$) are also used in a few experiments. Fluorescent Rhodamine B (Sigma) and a solvent-based color fluorescent (Kingscote-Mfr \#506250-RF4) are used to tag the silica nanoparticle dispersion and the oil phase, respectively. All products are used as received without further purification. \textbf{Note:} Unless otherwise stated, the silica dispersion, oil phase, and Span are refereed to the HS-40 Sigma silica dispersion, heptane, and Span 80 in the entire manuscript.\\ 
	\textbf{Sample preparation.}  We add DI-water (Direct-Q, Millipore Sigma) to the concentrated silica dispersion to reach the desired concentrations of 1.0, 2.0, 4.0, and 10.0 wt.\%. We prepare micellar solutions of Span in heptane with concentrations of 1.0, 2.5, 5.0, and 10.0 wt.\% and Span solutions with concentrations of 0.001, 0.01, 0.1, and 0.5 CMC. Span concentrations in mineral oils are fixed at 5.0 and 10.0 wt.\%. We use an ultrasonic bath (Isonic P4830) to homogenize all samples: 10 minutes for water and heptane based fluids and 60 minutes for the mineral oil based solutions. \\
	\textbf{Emulsion preparation.} We pour $3\ ml$ of an aqueous phase (DI-water or silica nanoparticle dispersions) into a glass vial  (1.25 oz) and add $3\ ml$ of an oil phase gently at the top of the aqueous phase using a micro-pipette. \\
	\textbf{Sample preparation for confocal microscopy.} We add Rhodamine B with a wavelength of 543 nm in the aqueous phase. Rhodamine B has a positive charge and tags negative particles (silica) in the water phase. The oil phase is dyed with a solvent-based fluorescent at the concentration of $2 \mu/10 m$l. We take a sample of emulsions using a micropipette, place it on a petri dish, and cover the sample with a microscopic glass slide to minimize the water and heptane evaporations during the imaging (5-10 minutes). \\ 
	\textbf{Confocal laser scanning microscopy.} Images are captured with 4X (CFI Plan Apo Lambda,
	Nikon) and 10X (CFI Plan Flour,Nikon) magnifications using confocal laser scanning microscopy (Nikon A1R). The laser power is set at 10-15 mW, and the pinhole is fixed at 1.2 AU for all images. We use 488 nm (FITC) and 561 nm (TRITC) laser emissions to detect the oil phase (green) and particles (red), respectively. \\
	\textbf{Cryo- Scanning electron microscopy.} We use a variable-pressure/environmental field emission scanning electron microscope (FEI Quanta 250 FEG) coupled with a Gatan Alto2500 cryotransfer system. Emulsion samples are freeze-dried in liquid Nitrogen before the imaging.\\
	\textbf{Interfacial tension of micellar solutions.} We measure the dynamic interfacial tension (IFT) of nanoparticle dispersions - micellar solutions using a spinning drop tensiometer (SDT, Kruss). We fill the glass capillary with the aqueous phase and place an oil drop at the top of its cap. The spinning rate is in the range of 6000-8000 rpm, and data are recorded for 4000 seconds. \\
	\textbf{Interfacial tension and elasticity measurements of Span solutions below CMC.} The dynamic IFT of nanoparticle dispersions - Span solutions at concentrations below CMC  and surface tension (ST) of the nanoparticle dispersions are measured using the pendant drop method (DSA100, Kruss). We form a pendant drop of the oil phase at a J-shaped stainless steel needle (OD=1.000 mm) immersed in a cuvette. The cuvette is filled with the silica nanoparticle dispersion. We record the IFT data for 2000 seconds. The elasticity of the interface is measured by imposing a sinusoidal perturbation in the surface area of the drop and analyzing the IFT response. We oscillate the drop with the frequency of 1 Hz and the amplitude of 5\% of the drop surface area, found through a systematic study, to ensure the linearity of the response.\\ 
	\textbf{Dynamic light scattering.} We measure the size of the reverse micelles and nanoparticles using dynamic light scattering (Malvern instrument, Nanoseries ZS). The average size is calculated according to $d\textsubscript{ave} = \sum d\textsubscript{i} n\textsubscript{i}$, where d\textsubscript{i} and n\textsubscript{i} are diameter and number percent of the reverse micelles.  \\
	\textbf{Density measurement.} We measure the density of the oil and aqueous phases before and after the oil-aqueous phases contact using Mettler Toledo densitometer (Density2Go).\\
	\textbf{Image analysis.} We use image Fiji software to analyze the images of emulsions and calculate the double emulsion intensity. Images are binarized with an appropriate threshold (80-100) value, and the number of droplets are counted using \textit{Analyze Particle} modulus.\\
	\textbf{Note.} All experiments are conducted at the atmospheric pressure ($\SI88.4 \pm0.8 kPa$) and temperature ($\SI21\pm0.5 ^{\circ} C$).  IFT, density, DLS experiments are repeated at least three times. Emulsion samples are generated at least ten times to confirm the reproducibility of the spontaneous double emulsion formation. Image analysis is repeated four times for each sample.\\
	\textbf{Data availability.} All raw data of DLS measurements are provided in graphs as supplemental materials.  All raw data for interfacial measurements and DLS will be provided in numerical forms in a spreadsheet upon request. 
	\newpage
	\newpage
	
	\newcounter{bean}
	\begin{list}%
		{[\arabic{bean}]}{\usecounter{bean}\setlength{\rightmargin}{\leftmargin}}
		
		\item G. Muschiolik, “Multiple emulsions for food use,” Current Opinion in Colloid \& Interface
		Science, vol. 12, no. 4-5, pp. 213–220, 2007.
		\item M. Hoppel, S. Juric, G. Reznicek, M. Wirth, and C. Valenta, “Multiple w/o/w emulsions
		as dermal peptide delivery systems,” Journal of Drug Delivery Science and Technology,
		vol. 25, pp. 16–22, 2015.
		\item J. Rodriguez, M. J. Martin, M. A. Ruiz, and B. Clares, “Current encapsulation strategies for bioactive oils: From alimentary to pharmaceutical perspectives,” Food Research
		International, vol. 83, pp. 41–59, 2016.
		\item J. Sunner, A. C. Calpena, B. Clares, C. Ca˜nadas, and L. Halbaut, “Development of clotrimazole multiple w/o/w emulsions as vehicles for drug delivery: effects of additives on
		emulsion stability,” AAPS PharmSciTech, vol. 18, no. 2, pp. 539–550, 2017.
		\item L. Ma, Z. Wan, and X. Yang, “Multiple water-in-oil-in-water emulsion gels based on selfassembled saponin fibrillar network for photosensitive cargo protection,” Journal of Agricultural and Food Chemistry, vol. 65, no. 44, pp. 9735–9743, 2017.
		\item A. Utada, E. Lorenceau, D. Link, P. Kaplan, H. A. Stone, and D. Weitz, “Monodisperse
		double emulsions generated from a microcapillary device,” Science, vol. 308, no. 5721,
		pp. 537–541, 2005.
		\item P. S. Clegg, J. W. Tavacoli, and P. J. Wilde, “One-step production of multiple emulsions:
		microfluidic, polymer-stabilized and particle-stabilized approaches,” Soft Matter, vol. 12,
		no. 4, pp. 998–1008, 2016.
		\item S. Ding, C. A. Serra, T. F. Vandamme, W. Yu, and N. Anton, “Double emulsions prepared by two–step emulsification: History, state-of-the-art and perspective,” Journal of
		Controlled Release, 2018.
		\item P. G. Moerman, P. C. Hohenberg, E. Vanden-Eijnden, and J. Brujic, “Emulsion patterns
		in the wake of a liquid–liquid phase separation front,” Proceedings of the National Academy
		of Sciences, vol. 115, no. 14, pp. 3599–3604, 2018.
		23
		\item S. Kim, K. Kim, and S. Q. Choi, “Controllable one-step double emulsion formation via
		phase inversion,” Soft Matter, vol. 14, no. 7, pp. 1094–1099, 2018.
		\item F. Sabri, W. Raphael, K. Berthomier, L. Fradette, J. R. Tavares, and N. Virgilio, “Onestep processing of highly viscous multiple pickering emulsions,” Journal of Colloid and
		Interface Science, 2019.
		\item C.-X. Zhao, D. Chen, Y. Hui, D. A. Weitz, and A. P. Middelberg, “Controlled generation
		of ultrathin-shell double emulsions and studies on their stability,” ChemPhysChem, vol. 18,
		no. 10, pp. 1393–1399, 2017.
		\item S. L. Anna, N. Bontoux, and H. A. Stone, “Formation of dispersions using “flow focusing”
		in microchannels,” Applied Physics Letters, vol. 82, no. 3, pp. 364–366, 2003.
		\item S. Takeuchi, P. Garstecki, D. B. Weibel, and G. M. Whitesides, “An axisymmetric flowfocusing microfluidic device,” Advanced Materials, vol. 17, no. 8, pp. 1067–1072, 2005.
		\item J. A. Hanson, C. B. Chang, S. M. Graves, Z. Li, T. G. Mason, and T. J. Deming, “Nanoscale
		double emulsions stabilized by single-component block copolypeptides,” Nature, vol. 455,
		no. 7209, p. 85, 2008.
		\item J. C. L\`{o}pez-Montilla, P. E. Herrera-Morales, S. Pandey, and D. O. Shah, “Spontaneous
		emulsification: mechanisms, physicochemical aspects, modeling, and applications,” Journal of Dispersion Science and Technology, vol. 23, no. 1-3, pp. 219–268, 2002.
		\item C. A. Miller, “Spontaneous emulsification produced by diffusion—a review,” Colloids and
		Surfaces, vol. 29, no. 1, pp. 89–102, 1988.
		\item B. F. Silva, C. Rodr\`{i}guez-Abreu, and N. Vilanova, “Recent advances in multiple emulsions and their application as templates,” Current Opinion in Colloid \& Interface Science,
		vol. 25, pp. 98–108, 2016.
		\item M. Schmitt, R. Toor, R. Denoyel, and M. Antoni, “Spontaneous microstructure formation
		at water/paraffin oil interfaces,” Langmuir, vol. 33, no. 49, pp. 14011–14019, 2017.
		\item Silva, P. S., Zhdanov, S., Starov, V. M., and Holdich, R. G. Spontaneous emulsification of water in oil at appreciable interfacial tensions. Colloids and Surfaces A: Physicochemical and Engineering Aspects vol. 521, no. 141-146, 2017.
		\item S. van der Graaf, C. Schr\"{o}en, and R. Boom, “Preparation of double emulsions by membrane
		emulsification—a review,” Journal of Membrane Science, vol. 251, no. 1-2, pp. 7–15, 2005.
		24
		\item W. D. Bancroft, “The theory of emulsification, v,” The Journal of Physical Chemistry,
		vol. 17, no. 6, pp. 501–519, 2002.
		\item J. Davies, “A quantitative kinetic theory of emulsion type. i. physical chemistry of the
		emulsifying agent,” in Gas/Liquid and Liquid/Liquid Interface. Proceedings of the International Congress of Surface Activity, vol. 1, pp. 426–438, 1957.
		\item B. P. Binks, “Particles as surfactants—similarities and differences,” Current Opinion in
		Colloid \& Interface Science, vol. 7, no. 1-2, pp. 21–41, 2002.
		\item B. P. Binks, R. Murakami, S. P. Armes, and S. Fujii, “Temperature-induced inversion
		of nanoparticle-stabilized emulsions,” Angewandte Chemie International Edition, vol. 44,
		no. 30, pp. 4795–4798, 2005.
		\item E. Read, S. Fujii, J. Amalvy, D. Randall, and S. Armes, “Effect of varying the oil phase on
		the behavior of ph-responsive latex-based emulsifiers: demulsification versus transitional
		phase inversion,” Langmuir, vol. 20, no. 18, pp. 7422–7429, 2004.
		\item B. P. Binks, J. Philip, and J. A. Rodrigues, “Inversion of silica-stabilized emulsions induced
		by particle concentration,” Langmuir, vol. 21, no. 8, pp. 3296–3302, 2005.
		\item M. Matos, A. Timgren, M. Sj\"{o}\"{o}, P. Dejmek, and M. Rayner, “Preparation and encapsulation properties of double pickering emulsions stabilized by quinoa starch granules,”
		Colloids and Surfaces A: Physicochemical and Engineering Aspects, vol. 423, pp. 147–153,
		2013.
		\item B. Binks and J. Rodrigues, “Types of phase inversion of silica particle stabilized emulsions
		containing triglyceride oil,” Langmuir, vol. 19, no. 12, pp. 4905–4912, 2003.
		\item B. P. Binks and C. P. Whitby, “Silica particle-stabilized emulsions of silicone oil and water:
		aspects of emulsification,” Langmuir, vol. 20, no. 4, pp. 1130–1137, 2004.
		\item F. Tu and D. Lee, “One-step encapsulation and triggered release based on janus particlestabilized multiple emulsions,” Chemical Communications, vol. 50, no. 98, pp. 15549–
		15552, 2014.
		25
		\item J. Bae, T. P. Russell, and R. C. Hayward, “Osmotically driven formation of double emulsions stabilized by amphiphilic block copolymers,” Angewandte Chemie International Edition, vol. 53, no. 31, pp. 8240–8245, 2014.
		\item C.-X. Zhao and A. P. Middelberg, “Microfluidic mass-transfer control for the simple formation of complex multiple emulsions,” Angewandte Chemie International Edition, vol. 48,
		no. 39, pp. 7208–7211, 2009.
		\item C.-H. Choi, D. A. Weitz, and C.-S. Lee, “One step formation of controllable complex
		emulsions: from functional particles to simultaneous encapsulation of hydrophilic and
		hydrophobic agents into desired position,” Advanced Materials, vol. 25, no. 18, pp. 2536–
		2541, 2013.
		\item S. K. Parida, S. Dash, S. Patel, and B. Mishra, “Adsorption of organic molecules on silica
		surface,” Advances in colloid and interface science, vol. 121, no. 1-3, pp. 77–110, 2006.
		\item F. Tiberg and T. Ederth, “Interfacial properties of nonionic surfactants and decane- surfactant microemulsions at the silica- water interface. an ellipsometry and surface force
		study,” The Journal of Physical Chemistry B, vol. 104, no. 41, pp. 9689–9695, 2000.
		\item H. Vatanparast, F. Shahabi, A. Bahramian, A. Javadi, and R. Miller, “The role of electrostatic repulsion on increasing surface activity of anionic surfactants in the presence of
		hydrophilic silica nanoparticles,” Scientific reports, vol. 8, no. 1, pp. 1–11, 2018.
		\item J. Jiang, Y. Zhu, Z. Cui, and B. P. Binks, “Switchable pickering emulsions stabilized
		by silica nanoparticles hydrophobized in situ with a switchable surfactant,” Angewandte
		Chemie International Edition, vol. 52, no. 47, pp. 12373–12376, 2013.
		\item F. Ravera, M. Ferrari, L. Liggieri, G. Loglio, E. Santini, and A. Zanobini, “Liquid–liquid
		interfacial properties of mixed nanoparticle–surfactant systems,” Colloids and Surfaces A:
		Physicochemical and Engineering Aspects, vol. 323, no. 1-3, pp. 99–108, 2008.
		
		\newpage
	\end{list}%

	{\large\bf SUPPLEMENTARY INFORMATION}
	\begin{center}
		
		{\large\bf Spontaneous Formation of Double Emulsions at Particle-Laden Interfaces}

		Parisa Bazazi and S. Hossein Hejazi
		
		{\it Department of Chemical and Petroleum Engineering, University of Calgary, Calgary, AB T2N 1N4,
			Canada}
	\end{center}

	\textbf{}\\
	\textbf{Supplementary Information}\\
	1. Spontaneous formation of emulsions
	\renewcommand{\thefigure}{S1}	
	\begin{figure}[H]
		\centering
		\includegraphics[scale=0.4]{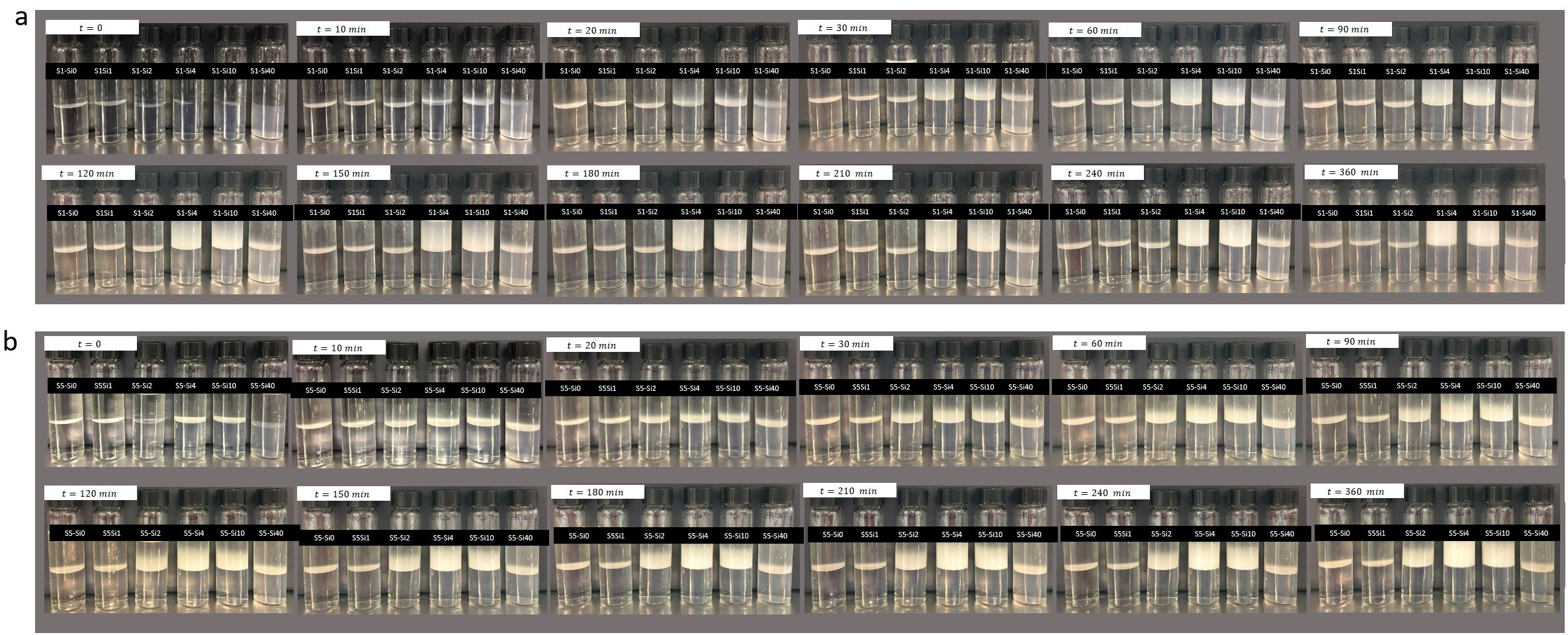}
		\caption{Formation of spontaneous emulsions in the bulk for \textbf{(a)} 1.0 wt.\%  and \textbf{(b)} 5.0 wt.\% Span micellar solutions in contact with 0.0, 1.0, 2.0, 4.0, 10.0, and 40.0 wt.\% Silica nanoparticle dispersions from left to right at different time scales. }
		\end{figure}
	\newpage
	\textbf{}\\
	2. Dynamic light scattering of micellar solutions 
		\renewcommand{\thefigure}{S2}
	\begin{figure}[H]
		\centering
		\includegraphics[scale=0.35]{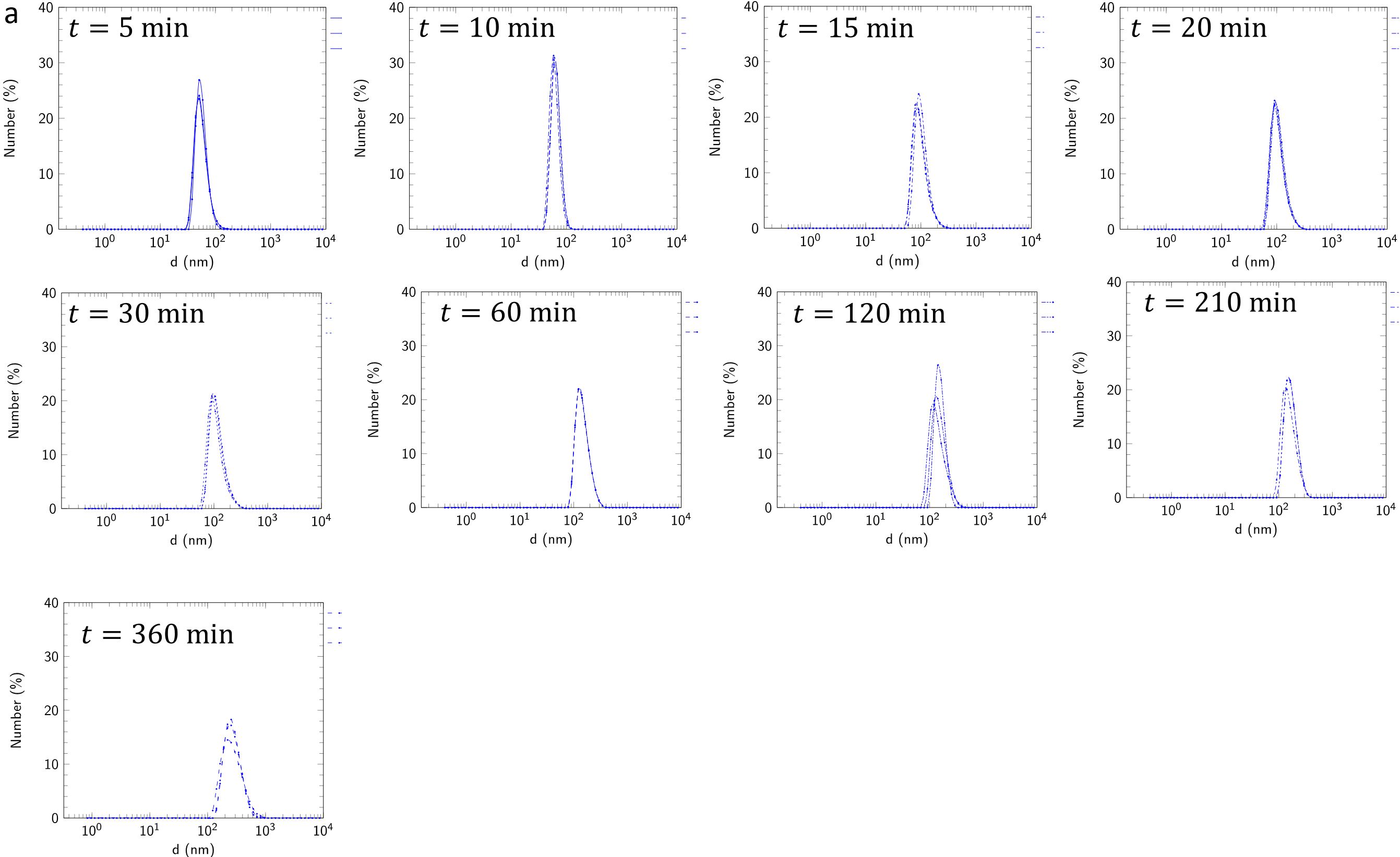}
		\caption{Size distribution of 1.0 wt.\%  Span micellar solutions in contact with 0.0 wt.\% Silica dispersion (DI-water).}
	\end{figure}
	\newpage
		\renewcommand{\thefigure}{S3}
	\begin{figure}[H]
		\centering
		\includegraphics[scale=0.35]{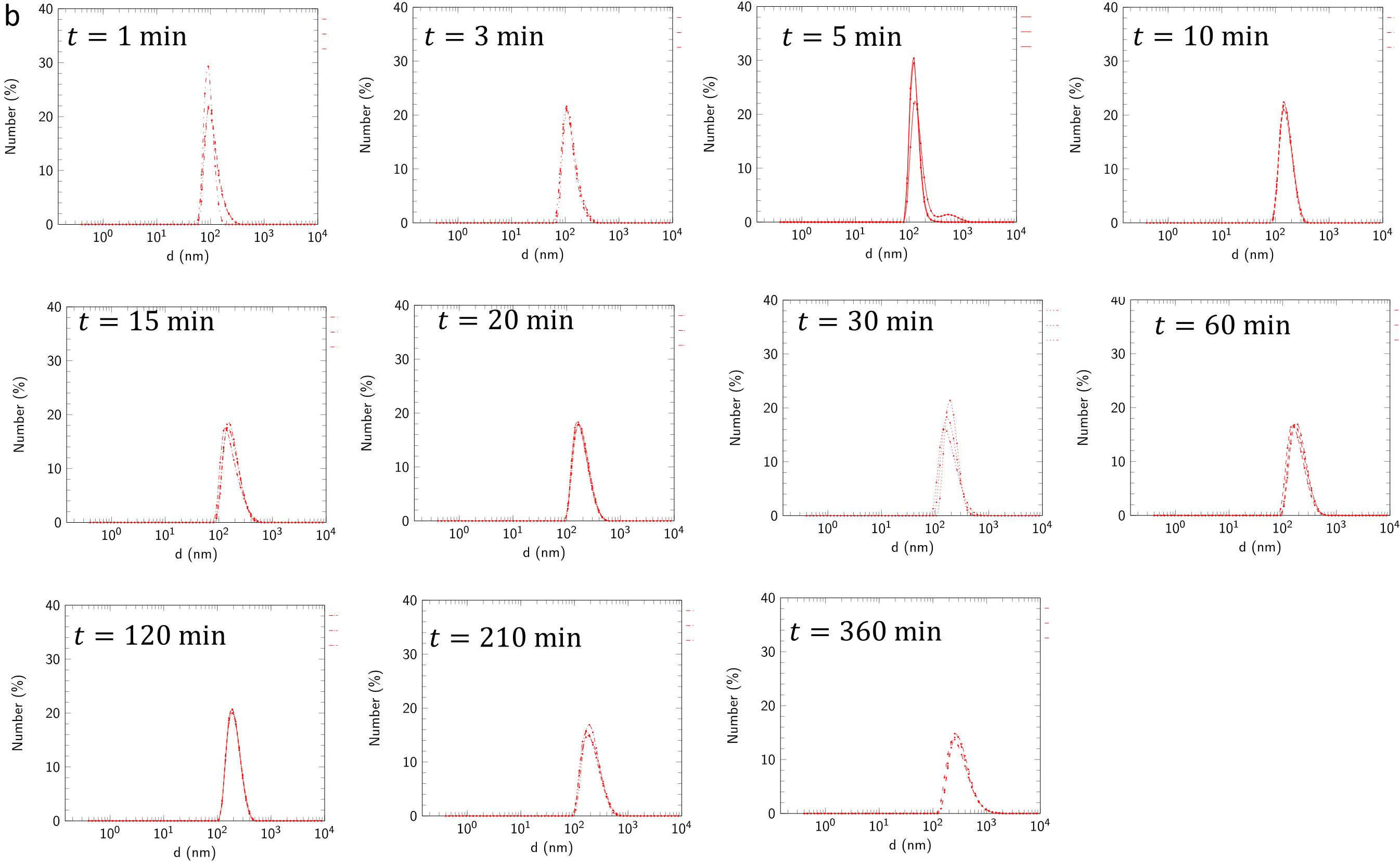}
		\caption{Size distribution of 1.0 wt.\% Span micellar solution in contact with 4.0 wt.\% Silica dispersion.}
	\end{figure}
	\newpage
		\renewcommand{\thefigure}{S4}
	\begin{figure}[H]
		\centering
		\includegraphics[scale=0.35]{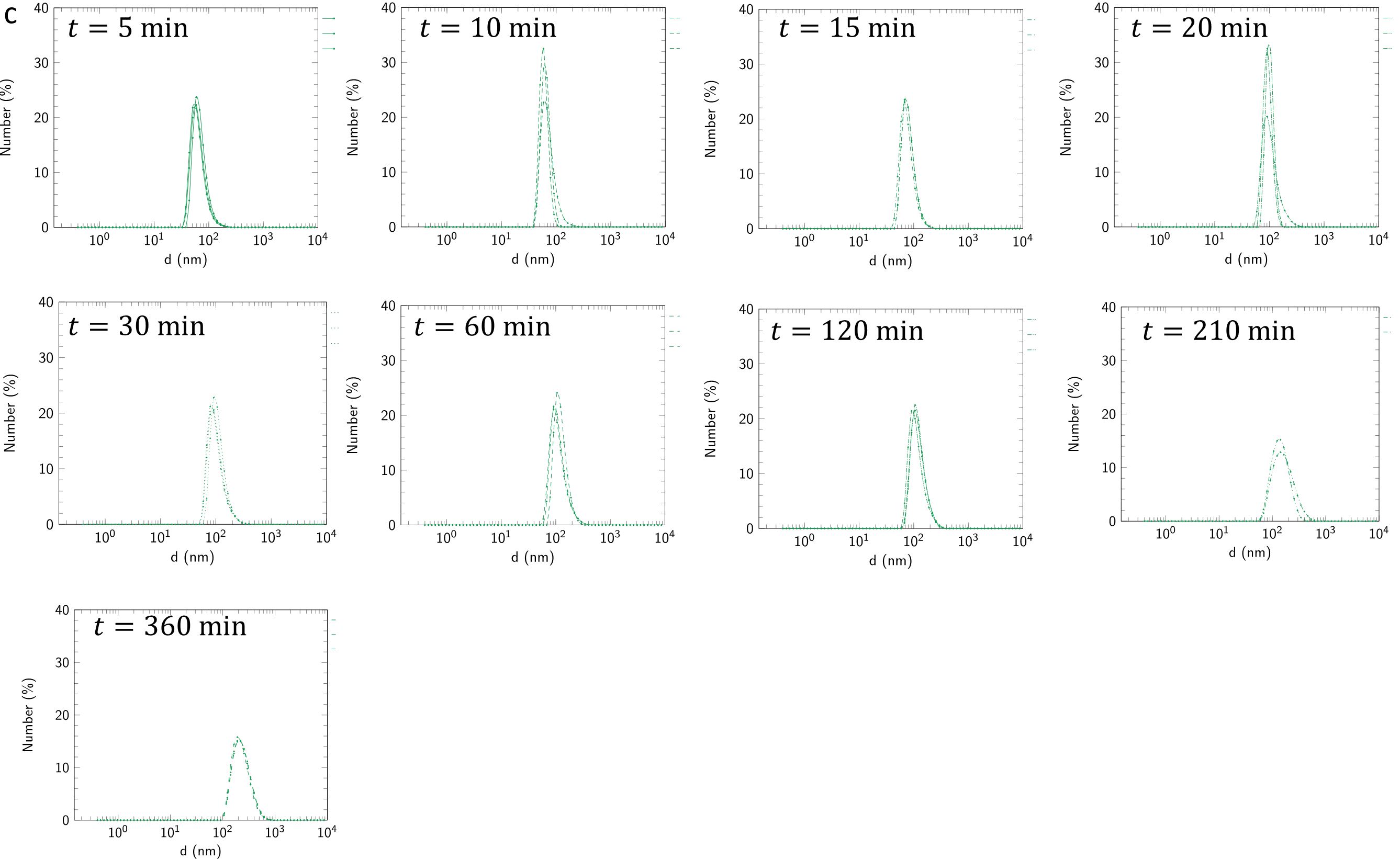}
		\caption{Size distribution of 5.0 wt.\% Span micellar solution in contact with 0.0 wt.\% Silica dispersion (DI-water).}
	\end{figure}
	\newpage
		\renewcommand{\thefigure}{S5}
	\begin{figure}[H]
		\centering
		\includegraphics[scale=0.35]{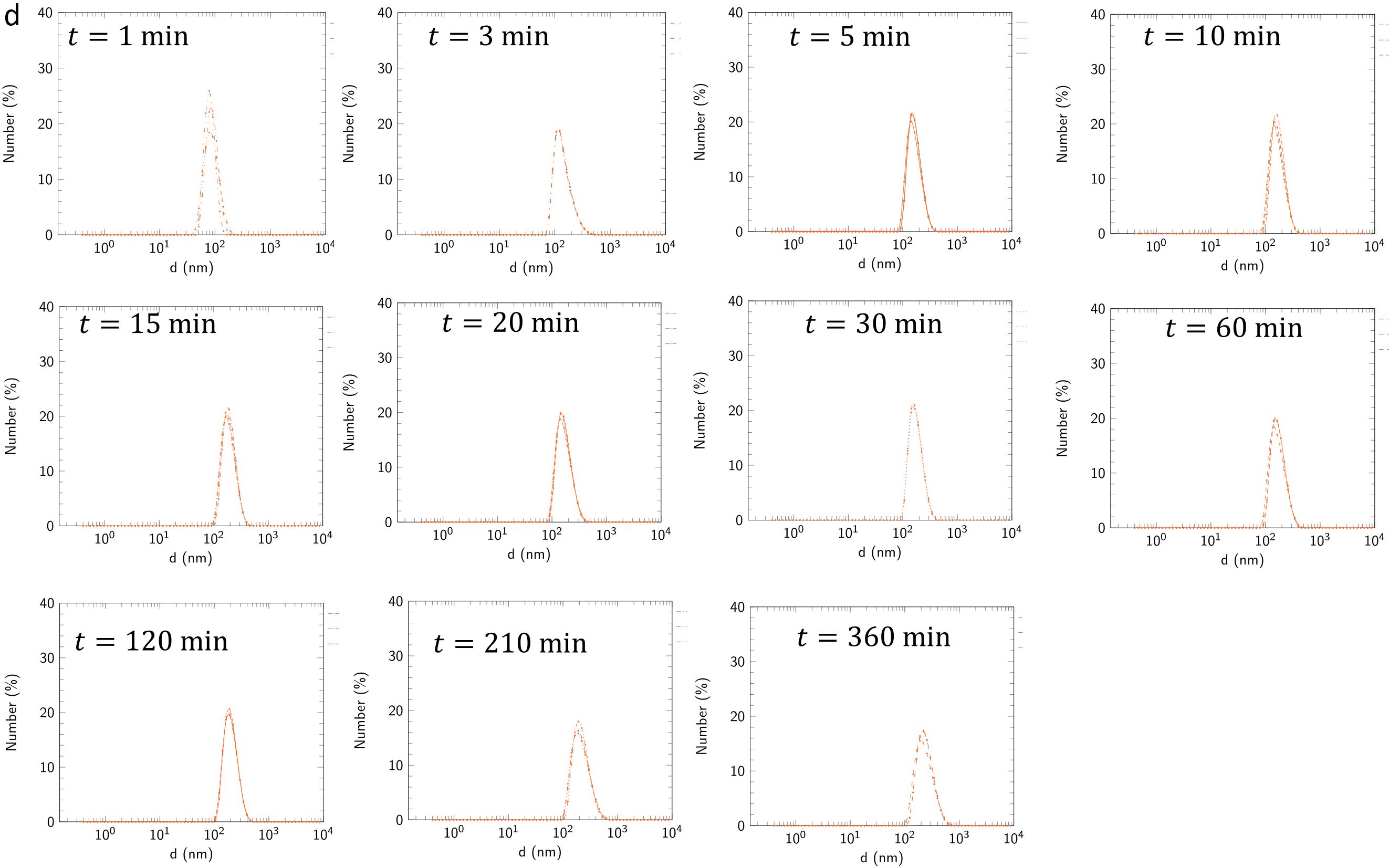}
		\caption{Size distribution of 5.0 wt.\% Span micellar solution in contact with 4.0 wt.\% Silica dispersion.}
	\end{figure}
	\newpage
	\textbf{}\\
	3. Confocal microscopy images of double emulsions over time
		\renewcommand{\thefigure}{S6}
	\begin{figure}[H]
		\centering
		\includegraphics[scale=0.35]{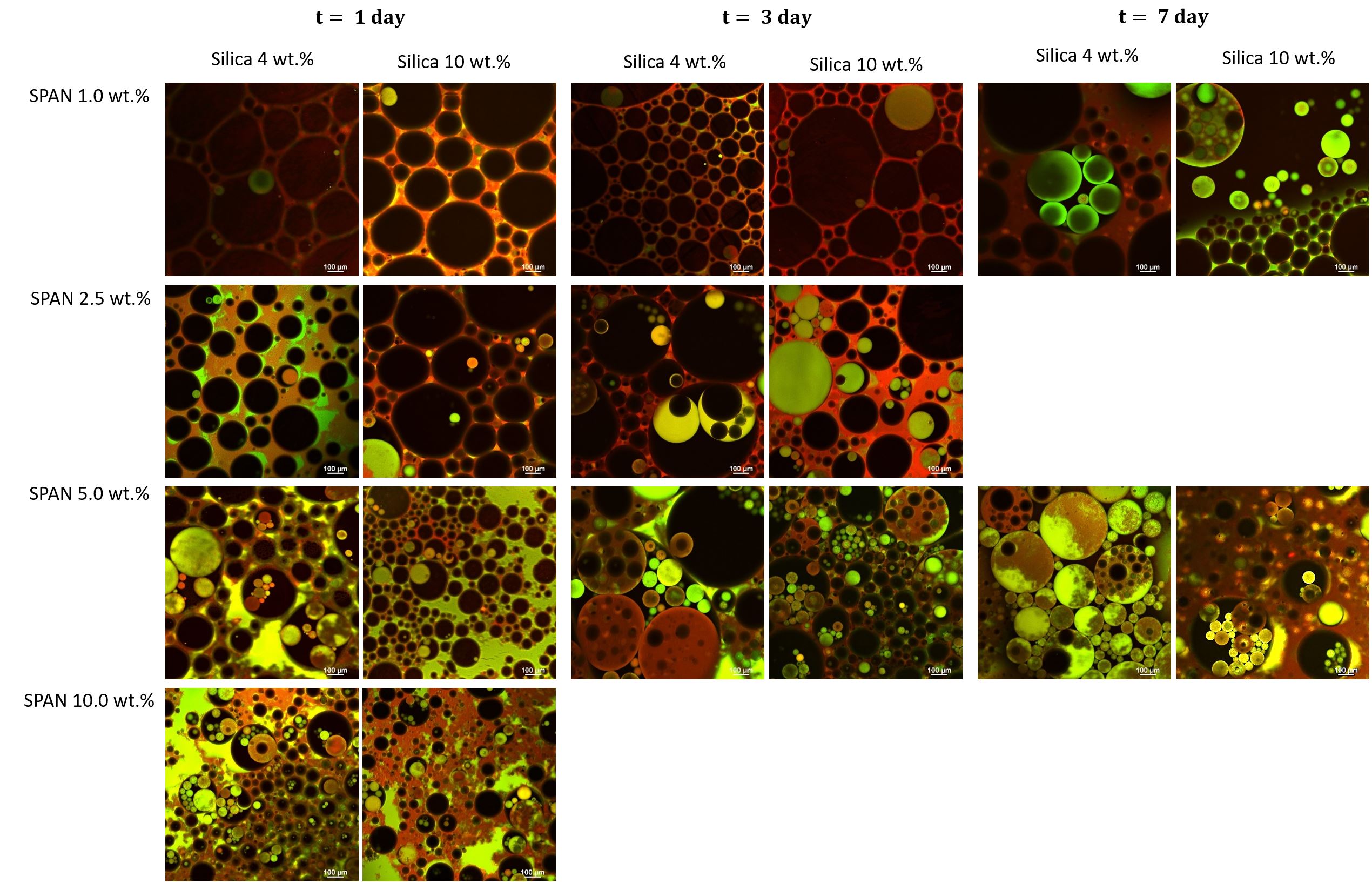}
		\caption{Confocal microscopy images of double emulsions with different concentrations of Span and Silica nanoparticle dispersions at 1, 3, and 7 days after the oil-aqueous phase contact. }
	\end{figure}
	\newpage
	\textbf{}\\
	4. Size distribution of nanoparticle dispersions and micellar solutions before the oil-aqueous phase contact, Density of 4.0 wt.\% Silica nanoparticle dispersion before and after the contact with micellar solutions.
		\renewcommand{\thefigure}{S7}
	\begin{figure}[H]
		\centering
		\includegraphics[scale=0.55]{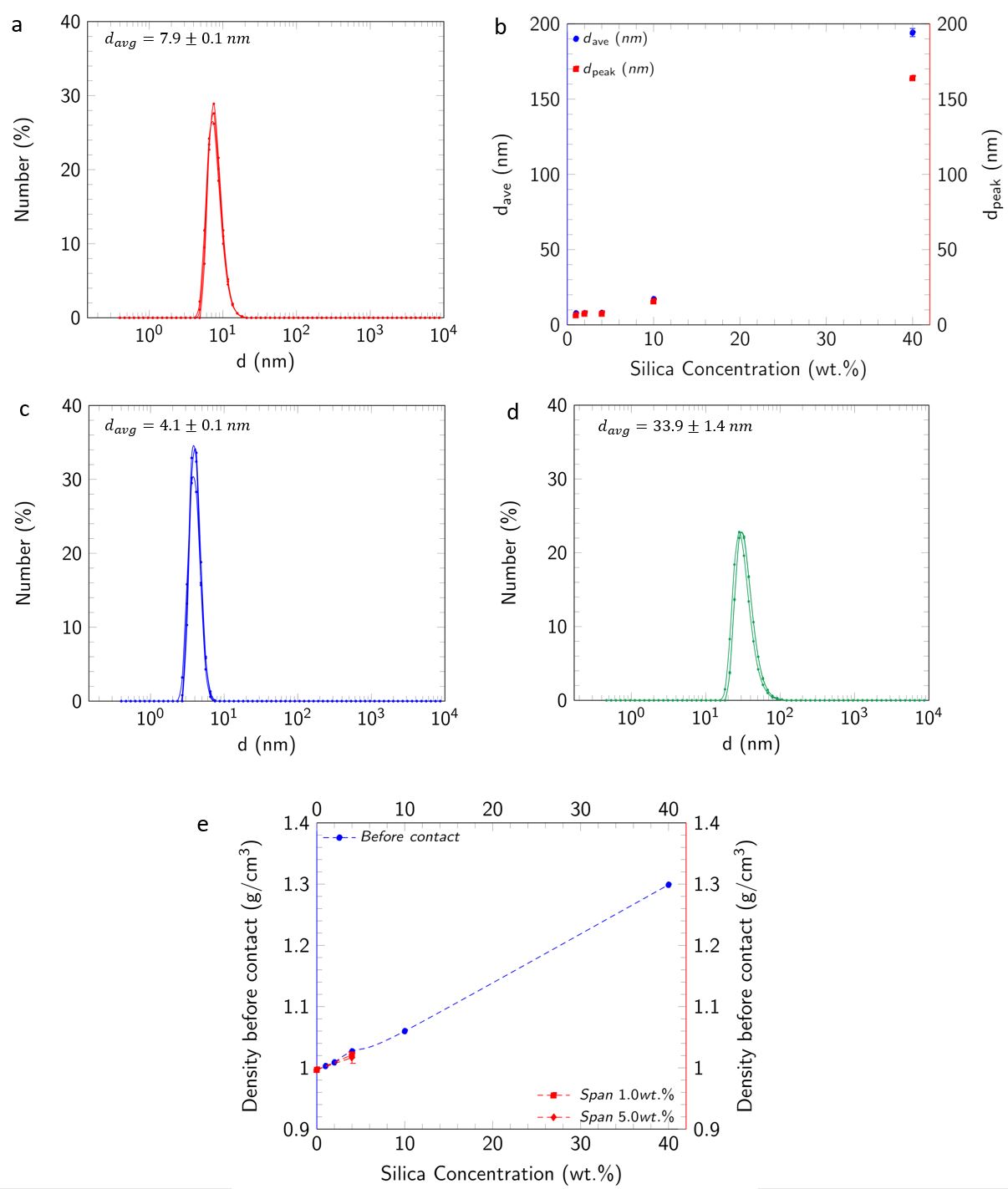}
		\caption{\textbf{(a)} Size distribution of nanoparticles in  4.0 wt.\% Silica nanoparticle dispersion. \textbf{(b)} Average size of Silica nanoparticles in 1.0, 2.0, 4.0, 10.0, and 40.0 wt.\% nanoparticle dispersions. \textbf{(c)} Size distribution of reverse micelles in 1.0 wt.\% Span micellar solution in heptane. \textbf{(d)} Size distribution of reverse micelles in 5.0 wt.\% Span micellar solution in heptane.(e) Density of the aqueous phase before contact (left axis, blue) and after the contact (right axis, red) with the 1.0 and 5.0 wt.\% Span micellar solutions.}
	\end{figure}
	\newpage
	
	\textbf{}\\
	5. Interfacial tension and interfacial elasticity of nanoparticle dispersions-below CMC Span solutions. 
		\renewcommand{\thefigure}{S8}
	\begin{figure}[h]
		\centering
		\includegraphics[scale=0.55]{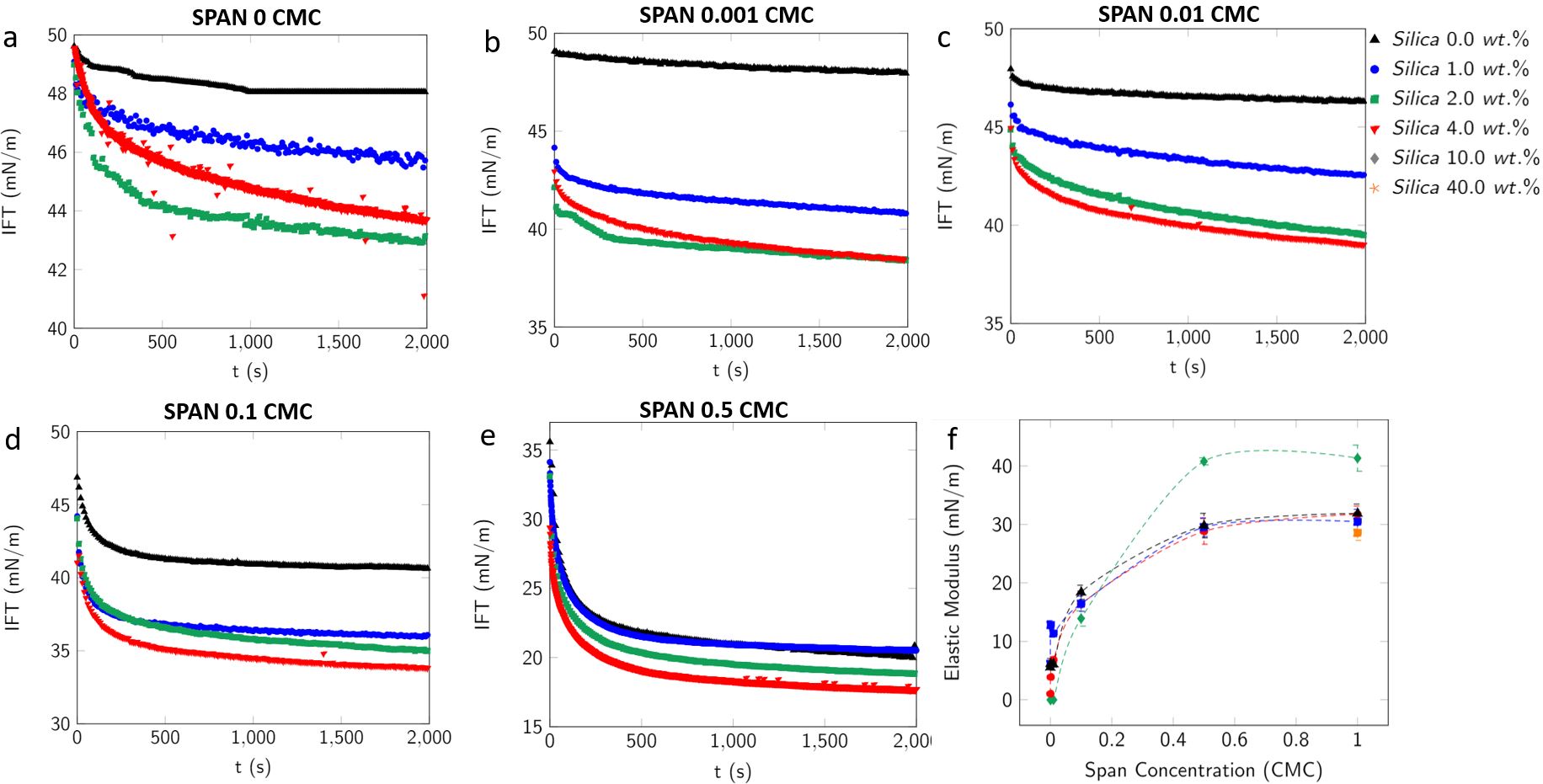}
		\caption{Interfacial tension of nanoparticle dispersions and Span 80 solutions of  0.000, 0.001, 0.010, 0.100, and 0.500 CMC in \textbf{a},\textbf{b}, \textbf{c}, \textbf{d}, and \textbf{e}. \textbf{e} Interfacial elasticity of nanoparticle dispersions with different Span concentrations.}
	\end{figure}
	
	\newpage
	\textbf{}\\
	6. Equilibrium and initial interfacial tension values of nanoparticle dispersion-Span 80 micellar solutions.
		\renewcommand{\thefigure}{S9}
	\begin{figure}[h]
		\centering
		\includegraphics[scale=0.55]{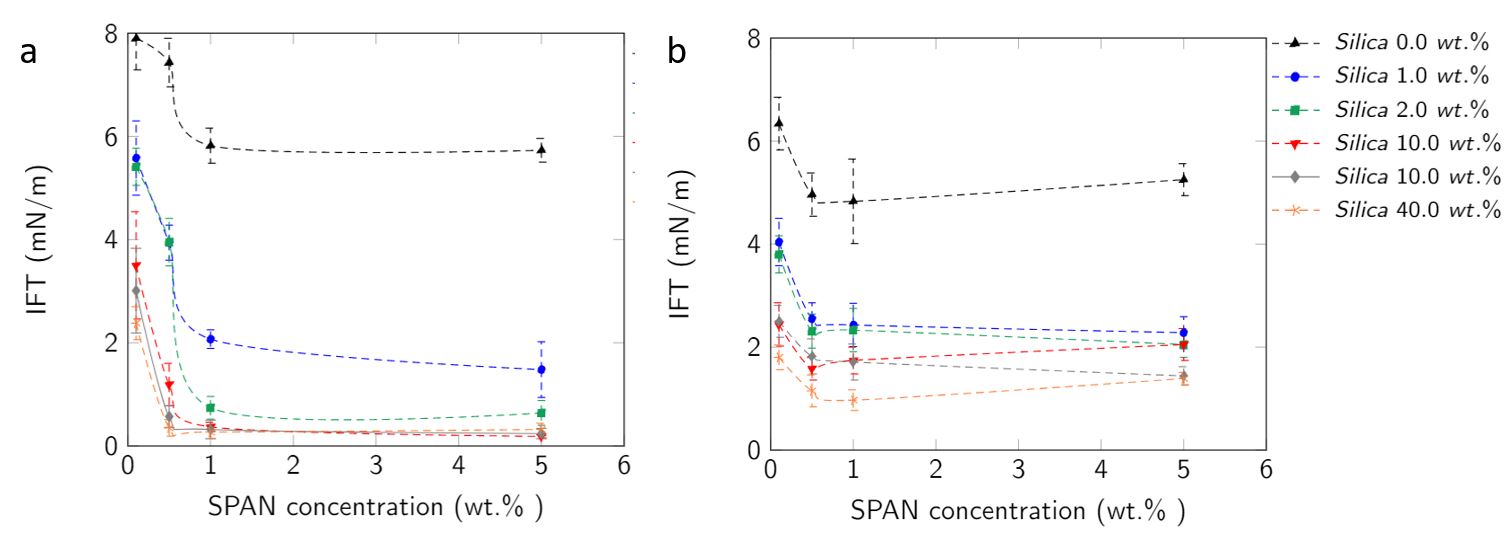}
		\caption{\textbf{(a)} Equilibrium and \textbf{(b)} initial interfacial tension values of nanoparticle dispersion-Span 80 micellar solutions}
	\end{figure}
	\newpage
	\textbf{}\\
	7. Spontaneous emulsion formation with mineral oils.
		\renewcommand{\thefigure}{S10}
	\begin{figure}[h]
		\centering
		\includegraphics[scale=0.65]{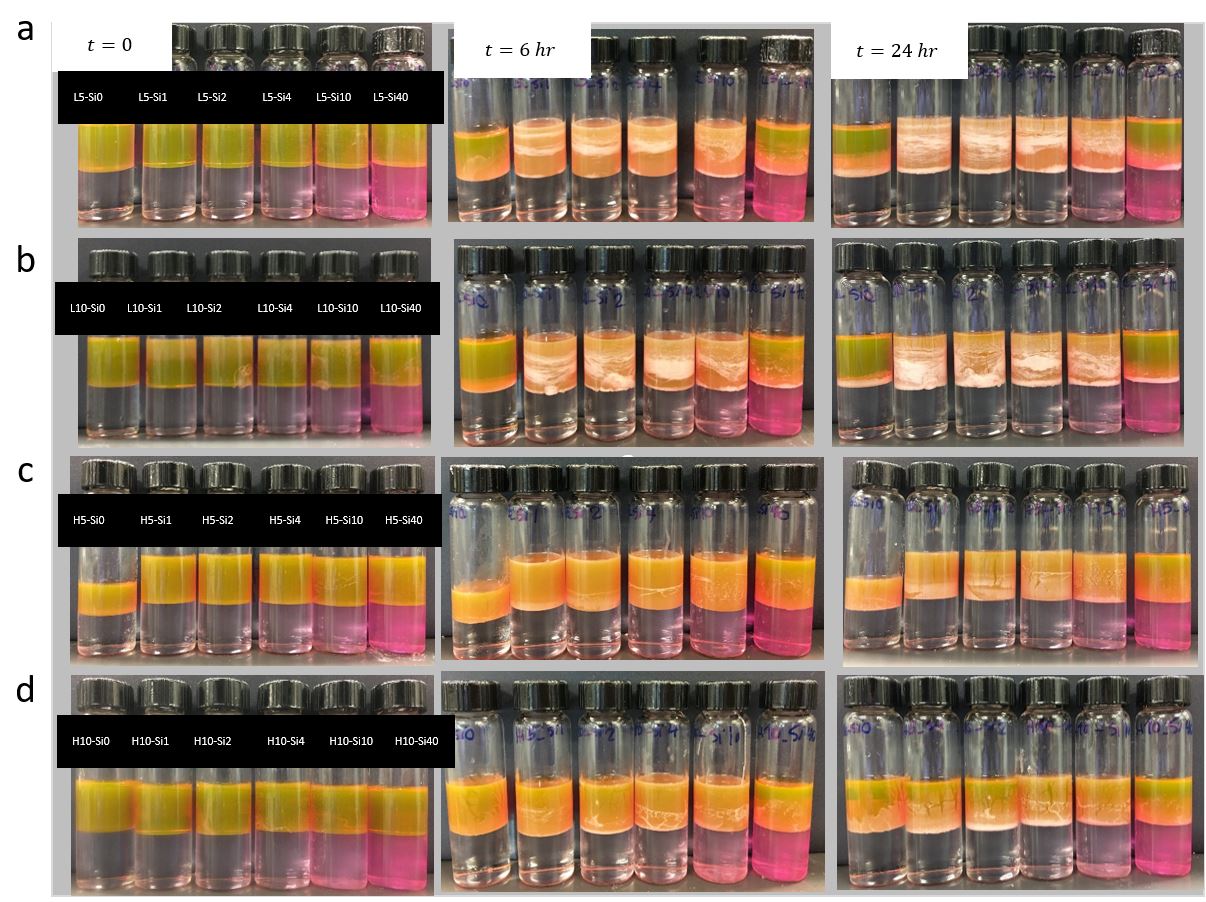}
		\caption{Formation of spontaneous emulsions in the bulk for \textbf{(a)} 5.0 wt.\%  Span in low viscosity mineral oil, \textbf{(b)} 10.0 wt.\% Span in low viscosity mineral oil,  \textbf{(c)} 5.0 wt.\%  Span in high viscosity mineral oil, \textbf{(d)} 10.0 wt.\% Span in high viscosity mineral oil in contact with 0.0, 1.0, 2.0, 4.0, 10.0, and 40.0 wt.\% Silica nanoparticle dispersions from left to right over time.}
	\end{figure}
	\newpage
	
	\newpage
\end{document}